\documentclass{IEEEtran}
\IEEEoverridecommandlockouts
\usepackage{cite}
\usepackage{amsmath,amssymb,amsfonts}
\usepackage{amsthm}
\usepackage{mathrsfs}  
\usepackage{algorithmic}
\usepackage{graphicx}
\usepackage{textcomp}
\usepackage{xcolor}
\usepackage{caption}
\usepackage{subcaption}
\def\BibTeX{{\rm B\kern-.05em{\sc i\kern-.025em b}\kern-.08em
    T\kern-.1667em\lower.7ex\hbox{E}\kern-.125emX}}

\usepackage{nomencl}
\usepackage{ifthen}
\usepackage{academicons}

\usepackage[latin1]{inputenc}

\usepackage{etoolbox}
\renewcommand\nomgroup[1]{%
  \item[\bfseries
  \ifstrequal{#1}{A}{Graph and Related Sets}{%
  \ifstrequal{#1}{B}{Signals and Instantaneous Quantities}{%
  \ifstrequal{#1}{C}{Constants}{}}}%
]}

\newtheorem{property}{Property}
\newtheorem{corollary}{Corollary}[property]
\newtheorem{application}{Application}[property]

\newcommand{\orcid}[1]{\href{https://orcid.org/#1}{\textcolor[HTML]{A6CE39}{\aiOrcid}}}
    
\begin{document}
\title{Characterizing the Effects of Single Bus Perturbation on Power Systems Graph Signals\\
\thanks{
This material is based upon work supported by the National Science Foundation under Grant  No.~2238658.
}
}

\author{Md Abul Hasnat, \IEEEmembership{Graduate Student Member, IEEE}, and Mia Naeini, \IEEEmembership{Senior Member, IEEE}

\thanks{Md A. Hasnat is with the Department of Electrical Engineering, University of South Florida, Tampa, 
FL 33620 USA (e-mail: hasnat@usf.edu).}
\thanks{Mia Naeini is with the Department of Electrical Engineering, University of South Florida, Tampa, 
FL 33620 USA (e-mail: mnaeini@usf.edu).}
\thanks{Corresponding author: Md Abul Hasnat.}}

\maketitle

\begin{abstract}
This article explores the effects of a single bus perturbation in the electrical grid using a Graph Signal Processing (GSP) perspective. The perturbation is characterized by a sudden change in real-power load demand or generation. The study focuses on analyzing the spread of the perturbation throughout the grid and proposes a measure of spreadability based on GSP.  Moreover, the global and local smoothness properties of the difference bus voltage angle graph signals are evaluated for understanding their embedded patterns of spreadability property. It is demonstrated that the global smoothness of the bus voltage angle graph signal follows a quadratic relationship with the perturbation strength, which helps in characterizing the critical perturbation strength after which the power flow diverges indicating a stressed system. The impact of a single bus perturbation on power system graph signals has been investigated through both analytical derivations using the DC power flow model and simulation using the AC power flow model. The results reveal that the proposed measure of spreadability as well as local and global smoothness properties of the graph signals are independent of the perturbation strength and instead mainly depend on the perturbation's location.
\end{abstract}


\makenomenclature

\nomenclature[A, 01]{$\mathcal{V}$}{Set of all buses (vertices).}
\nomenclature[A, 01]{$\mathcal{S}$}{Set of all generator buses.}
\nomenclature[A, 01]{$\mathcal{L}$}{Set of all load buses.}

\nomenclature[A, 01]{$\mathcal{G}$}{Graph associated with the power system (i.e., the domain of the graph signal).}
\nomenclature[A, 01]{$\mathcal{G}^{\prime}$}{Unweighted version of $\mathcal{G}$.}
\nomenclature[A, 01]{$\mathcal{E}$}{Set of all transmission lines (edges).}
\nomenclature[A, 01]{$\mathcal{W}$}{Set of all edge weights.}
\nomenclature[A, 01]{${\mathcal{N}}_u^{(K)}$}{Set of the $K-$ hop neighbors of $v_u$.}
\nomenclature[A, 01]{$\mathscr{S}$}{Slope operator.}
\nomenclature[A, 01]{$\mathscr{D}(v_i,v_j)$}{Shortest path distance operator between vertices $v_i$ and $v_j$.}

\nomenclature[A, 01]{$\mathbf{L}$}{Graph Laplacian Matrix.}
\nomenclature[A, 01]{$\mathbf{B}$}{Susceptance Matrix.}
\nomenclature[A, 01]{$\mathbf{Q}$}{{Matrix containing information about grid topology and electrical distances defined as \small{$\left(\mathbf{B}^{-1}\right)^{T} \mathbf{L} \mathbf{B}^{-1}$}}.} 
\nomenclature[A, 01]{$\mathbf{R}$}{{Matrix containing information about grid topology and electrical distances defined as \small{$\left(\mathbf{B}^{-1}\right)^{T} \mathbf{B}^{-1}$}}.}

\nomenclature[A, 01]{$d_{ij}$}{Geographical distance between bus $i$ and $j$.}

\nomenclature[A, 02]{$e_{ij}$}{Link between vertex $v_i$ and $v_j$.}

\nomenclature[A, 02]{$w_{ij}$}{Weight of the link $e_{ij}$.}

\nomenclature[A, 02]{$l_{ij}$}{Entry of row $i$ and column $j$ of $\mathbf{L}$.}
\nomenclature[A, 02]{$b_{ij}$}{Entry of row $i$ and column $j$ of $\mathbf{B}$.}
\nomenclature[A, 02]{$\beta_{ij}$}{Entry of row $i$ and column $j$ of ${\mathbf{B}}^{-1}$.}

\nomenclature[A, 06]{$\lambda_k$}{$k-$th eigenvalue of $\mathbf{L}$.}

\nomenclature[B, 01]{$x(v_n), x(n)$}{Graph signal in general.}
\nomenclature[B, 01]{$x(n,t)$}{Time-varying graph signal.}
\nomenclature[B, 01]{$p(n)$}{Bus real power graph Signal.}
\nomenclature[B, 01]{$p_d(n)$}{Actual load demand graph signal.}
\nomenclature[B, 01]{$p_d(n)$}{Generated real power graph signal.}
\nomenclature[B, 01]{$\underline{\mathbf{x}}$}{Graph signal $x(n)$ in vector form.}
\nomenclature[B, 01]{$\theta(n)$}{Bus voltage angle graph signal.}
\nomenclature[B, 01]{$\Delta \theta(n)$}{Difference bus voltage angle graph signal, after and before the perturbation.}
\nomenclature[B, 01]{$\psi_u(n)$}{Normalized difference voltage angle graph signal.}
\nomenclature[B, 01]{$g_x$}{Global smoothness of graph signal $x(n)$.}
\nomenclature[B, 01]{$l_x(n)$}{Local smoothness of graph signal $x(n)$.}

\nomenclature[B, 01]{$C^{\prime}(n)$}{Modified normalized closeness centrality of vertex $v_n$.}

\nomenclature[B, 01]{$f_{y}(\zeta)$}{Probability distribution of random variable $y$.}

\nomenclature[C, 01]{$N$}{Cardinality of the set $\mathcal{V}$.}

\nomenclature[C, 01]{$t_{\text{u}}$}{Perturbation instant.}
\nomenclature[C, 01]{${\bar{\psi}}_u^{(K)}$}{Mean of the signal values of $\psi(n)$ for all the vertices at $K-$ hop distance from the perturbed bus.}
\nomenclature[C, 01]{$\gamma$}{Perturbation strength.}
\nomenclature[C, 01]{$\gamma_c$}{Critical perturbation strength.}
\nomenclature[C, 01]{$\gamma_{nc}$}{Non-convergence perturbation strength.}
\nomenclature[C, 01]{$K$}{Hop distance.}

\printnomenclature[1.8cm]

\section{Introduction}
Graph signal processing (GSP) has emerged as a prominent field that focuses on the analysis of structured data over the graph domain. Recently, GSP has found applications in the analysis of power system data by representing the power system as a graph and its measurements over the graph as graph signals \cite{ramakrishna21, hasnat22}. By extending the theories and tools of classical signal processing to the irregular graph domain, GSP facilitates imparting explicit information about the topology, connectivity, and interactions among the components of the system into the analysis of data. Detection, localization, and classification of anomalies, attacks, and stresses in the electric grid \cite{ramakrishna21, hasnat22, takiddin23}, state estimation and recovery \cite{saha22, hasnat22pesgm, dabush23},  estimation of load current variability in the presence of distributed generators \cite{mendes23}, and load disaggregation \cite{he18} are examples of applications of GSP in addressing problems in power systems. 

Analyzing power grid data through the lens of GSP has revealed that signatures and patterns of stresses in the system are embedded in various properties and features related to the system's graph signals \cite{hasnat22,hasnat21isgt}. In this work, the focus is on understanding the features and patterns in power systems graph signals due to abrupt changes in the load demand or generated power in a single bus. Although fluctuation of load demand within an acceptable range is normal and perpetual in the power system, understanding the patterns of load change is important for situational awareness, particularly in the context of smart grids with intermittent and low-inertia loads \cite{ratnam20}. A typical scenario is the charging of electrical vehicles (EVs) as a load added to the grid (G2V technology) \cite{jain14}. Since the load demand associated with the charging of the EVs is more probable to be clustered geographically \cite{mullan11}, a monotonous increase of load demand at a particular bus can be a common situation. Another origin of the monotonous increase in load demand can be the load-altering cyber attacks purposefully launched by adversaries \cite{stamp09,mohsenian-rad11}. The abrupt changes in the generation of real power are not common in traditional power systems but are possible in modern power grids when a large number of renewable energy resources are connected to the grid by converters \cite{ratnam20}.  In this work, a general approach has been considered, from the GSP perspective, to analyze the effects of changes in the load demand or generated real power at a particular bus, modeled as {\it single bus perturbation}, without explicitly modeling the cause of perturbation. 

The first presented study is focused on understanding how a single bus perturbation spreads through the power grid depending on the strength and the location of the perturbation. The analysis of the spreadability of a bus perturbation is important from several perspectives in the context of grid stability and reliability analysis. A more spreadable perturbation can affect a large number of components (e.g., buses, transmission lines), even at distant locations from the perturbation point, and introduces stresses in the grid that may even lead to cascading failures or blackouts. Here, a GSP-based measure is defined to quantify the spreadability of the perturbation depending on its strength and location. This spreadability measure is also useful for planning the placement of low-inertia loads and generators in the grid.

In addition to understanding the spreadability, it is important to understand how the perturbation affects other graph signal features to gain an improved situational awareness under stress. For instance, power system graph signals, especially the bus voltage angle graph signal, are generally smooth during normal grid operation \cite{ramakrishna21, hasnat22, hasnat21isgt}; however, the local and global smoothness properties of the graph signals vary under stress. This study focuses on understanding how the global and local smoothness values associated with the power system graph signals are affected as a function of the perturbation strength and location. The relation between the proposed spreadability measure and local and global smoothness features of graph signals has also been explored and it has been shown that certain smoothness parameters associated with the difference graph signals (before and after perturbation) can be good estimators of the spreadability of the perturbations. 

The effects of single bus perturbation on the power system graph signals have been derived analytically using the DC power flow model and simulated using the AC power flow model to verify the properties in more realistic scenarios. The presented analytical approach shows that the proposed measure of spreadability does not depend on the perturbation strength, but rather depends on the location of the perturbation. Our experiment based on the AC power flow model closely supports this property. Moreover, the presented analytical analysis shows that the global smoothness of the bus voltage angle graph signal is a quadratic function of the increasing load demand (or generated real power) at a particular bus. Based on this analysis, there is a critical value of input power at each bus beyond which the global smoothness begins to drop, and a further increase in the input power leads to divergence of the power flow equations. Failing of power flow convergence, although arises from various issues, is an indicator of a stressed system. The presented analytical study shows that the critical load (or generation) at each bus for which the global smoothness is maximized depends on the topology.

The key contributions of this article have been summarized below:
\begin{itemize}
    \item A quantitative measure of the spreadability of a perturbation has been proposed and the properties of this measure have been analyzed theoretically under the DC power flow model and verified using the AC power flow results. The proposed measure has been compared with an existing network-science-based spreadability metric.  

    \item The global smoothness of the voltage angle graph signal has been shown to follow a quadratic function of the perturbation strength with a maximum, defined as the critical perturbation. It is shown that this critical point suggests approaching the power flow model divergence, which although can arise from various issues, is an indicator of a stressed grid.

    \item 
    The global and local smoothness properties of the difference graph signal of bus voltage angles before and after the perturbation are examined. 
    The analysis demonstrated that under DC plow flow assumptions these smoothness parameters are independent of the perturbation strength and thus are suitable for analyzing the effects of perturbations at different locations of the grid. The results of the simulation with the AC power flow model closely support this property. Moreover, these smoothness parameters have been identified as reliable indicators of the spreadability of perturbations based on the location.  
    
\end{itemize}

\section{Related Work}

The effects of perturbations in the electrical grid have been studied from various perspectives in the literature. The stability of the grid after the perturbation, the dependency on the perturbation location, the propagation of the effect of perturbation through the system, and the identification of vulnerable locations in the grid are some of the topics of interest in this domain. A number of works analyze the effects of perturbation from the complex network perspective using the concept of \textit{Basin stability} \cite{wolff18, menck12} using the frequency measurements in the grid. For example, Wolff \textit{et. al.} \cite{wolff18} analyzed the effect of perturbation of a single node (bus) in the electric grid based on the Basin stability of the grid, which is evaluated in terms of the return time of the grid to the steady-state after the perturbation. This work defines perturbation as the direct change of voltage phase angle and angular frequency at the perturbed bus. Menck and Kurths \cite{menck12} identify the weaker buses due to small perturbations in the grid based on Basin stability. 

The propagation of spatio-temporal signals through the system has been studied by several authors with a complex network approach. Hens \textit{et. al.} \cite{hens19} provide a generalized theoretical analysis of how spatio-temporal signals propagate in time through complex networks depending on the topology and dynamic mechanisms of interactions among the vertices. A few works also studied the spread of disturbances in the electric power grid. For example, Molner \textit{et. al.} \cite{molnar21} proposed a heuristic technique to relate the spread of oscillations due to the variable renewable resources to the network structure. Nnoli and Kettemann \cite{nnoli21} analyzed the propagation of disturbance in the electric grid depending on the topology of the grid, its inertia, and heterogeneity. In \cite{buttner22}, the authors considered a network-science-based approach to quantify the spreadability of a single perturbation in the grid depending on the perturbation location.
  
The impact analysis of grid perturbations can be useful in several scenarios in the modern power grids including the integration of distributed energy resources (DERs) and electric vehicle charging stations. Although in most cases, the problems do not directly correspond to the single bus perturbation, the single perturbation analysis can be useful for the simplification of such problems. In the current literature, the issues related to the integration of EVs and DERs have been studied using various methods. For instance, Vasilij \textit{et. al.} \cite{vasilj22} developed a model for the worst-case analysis of the impact of placing EV charging stations in the grid, which involves observing the impact of the placement of charging stations on voltage profile and line loading. 

The current work presents a generalized approach to analyze the impact of a single bus perturbation in the grid. Moreover, unlike the Basin stability-based analyses, this work does not consider the frequency data and only considers the impact of the perturbation on the bus voltage angle data. The current work adds a GSP perspective to the analysis to directly impart the topology and interconnection into the analysis.  

\section{Mathematical Representation of Perturbation and Associated Electrical Attributes} 

\subsection{Power System Graph Signals}

An electric power grid with $N$ buses and $M$ transmission lines has been modeled as a weighted undirected graph, $\mathcal{G}=(\mathcal{V},\mathcal{E}, \mathcal{W})$. The buses of the grid are considered as the vertices of the $\mathcal{V}=\{v_1, v_2, ..., v_N\}$, whereas the transmission lines are considered as the edges, $\mathcal{E}=\{e_{ij}: (i,j) \in \mathcal{V} \times \mathcal{V}\}$, and therefore, $|\mathcal{V}|=N$ and $|\mathcal{E}|=M$, where $|.|$ denotes the cardinality of the set. The element $w_{ij}$ of the weight matrix, $\mathcal{W}$ is the weight corresponding to the edge, $e_{ij}$. The vertices corresponding to the buses with generators (i.e. energy sources) and loads are denoted by $\mathcal{S} \subset \mathcal{V}$ and $\mathcal{L} \subset \mathcal{V}$, respectively. The Laplacian matrix $\mathbf{L}$ associated with the graph $\mathcal{G}$ with elements $l_{ij}$ is defined as: $l_{ij}=\sum_{j=1}^N w_{ij}$, if $i=j$ and $l_{ij}=-w_{ij}$, otherwise. In the GSP literature, the weights are defined in various ways, for instance, based on the geographical and physical relational aspects, depending on the applications. In this work, the weights $w_{ij}$ are defined such that the Laplacian matrix, $\mathbf{L}$ represents the imaginary part of the admittance matrix of the grid capturing some of the transmission line properties.  

The graph signal $x(v_n)$, written as $x(n)$ for simplicity, can be considered as a mapping of the vertices of the graph to real-number space, $x:\mathcal{V} \rightarrow \mathbb{R}$ and can represent various electrical attributes associated with the buses of the grid. The signal values of $x(n)$ arranged in a vector form would be denoted by $\underline{\mathbf{x}}$. In this article, the graph signal $x(n)$ at a particular time instant $t$ is denoted as $x(n,t)$. 

Let us consider $\theta(n)$, the bus voltage angle graph signal that represents the angles of the voltage phasors at each bus. While any or combination of electrical attributes at each bus can be considered, here the focus will be on the voltage angle graph signal $\theta(n)$ to evaluate the state of the power system without the direct information on the transient state based on the fluctuations in the voltage magnitudes and frequency. Moreover, bus voltage angle measurements are directly related to the load demands, which are important in this study. The generated real power and the real power demand at each bus are denoted by the \textit{generated power graph signal}, $p_g(n)$ and \textit{load demand graph signal}, $p_d(n)$, respectively. Note that $p_g(n)=0$ for $n \in \mathcal{V} \setminus \mathcal{S}$ and $p_d(n)=0$ for $n \in \mathcal{V} \setminus \mathcal{L}$. The \textit{input power graph signal} is denoted by $p(n)$, where $p(n)=p_g(n)-p_d(n)$.

\subsection{DC Power Flow Model}
The DC power flow model \cite{zimmerman11} describes a linear relationship between the input power and the bus voltage angle by the equation ${\underline{\mathbf{p}}} = \mathbf{B} {\underline{\mathbf{\theta}}}$, where $\mathbf{B}$ is the susceptance matrix of the grid (imaginary part of the admittance matrix) with element $b_{ij}$ at the $i-$th row and $j-$th column. Knowing the topology, the bus voltages can be computed from the active power input based on ${\underline{\mathbf{\theta}}}=\mathbf{B^{-1}} \underline{\mathbf{p}}$ and can be represented in a graph signal form as:
\begin{equation} \label{dcpf}
    \theta(n) = \sum_{j=1}^{N} {\beta}_{nj} p(j),
\end{equation}
where ${\beta}_{ij}$ is the element of $\mathbf{B^{-1}}$ at the $i-$th row and the $j-$th column. In this work, the linearity of the DC power flow model facilitates analytical investigation of the properties of the graph signals. However, since the DC power flow model is an approximation of the power flow in power systems, in certain cases, the results from this model may deviate from the real scenarios. Nevertheless, the graph signal analysis with the DC power flow assumption reveals important information about the state of the system. Whenever necessary, in this work, the AC power flow model through MATPOWER \cite{zimmerman11} is utilized for numerical verification of the analytical results. 

\subsection{Smoothness of Graph Signals} 

The global smoothness of a graph signal is a measurement of the overall amount of vertex-to-vertex fluctuations in the graph signal \cite{dakovic19}. The global smoothness value associated with graph signal, $x(n)$ is defined as \cite{dakovic19}: 
\begin{equation}\label{defg}
    g_x = \frac{{\underline{\mathbf{x}}}^T \mathbf{L} \underline{\mathbf{x}}}{{\underline{\mathbf{x}}}^T  \underline{\mathbf{x}}} = \frac{\sum_{i=1}^N \sum_{j=1}^N L_{ij} x(i) x(j)}{\sum_{k=1}^N x^2(k)}.
\end{equation}
A small value of $g_x$ indicates a smooth graph signal, whereas increasing values of $g_x$ indicate the increasing vertex-to-vertex fluctuations of signal values \cite{dakovic19}. The bus voltage angle graph signal $\theta(n)$ in normal conditions is generally smooth with a small value of $g_{\theta}$ \cite{hasnat22}. 

The local smoothness \cite{dakovic19} of a graph signal $x(n)$ is defined by the following equation and represents how rapidly the value of a graph signal changes from each vertex $n$ to its neighboring vertices: 
\begin{equation}\label{deflocsmooth}
    l_{x}(n)= \frac{\sum_{k=1}^N L_{nk} x(k)}{x(n)}, \ \  x(n) \neq 0. 
\end{equation}

Our previous analyses of the local smoothness for the bus voltage angle graph signals in power systems, in \cite{hasnat22, hasnat22a}, have revealed that the voltage angle graph signals are smoother at certain locations in the grid depending on the topology and interconnections among the components of the system. 

The global and local smoothness values of graph signals are important features in the vertex domain that can allow analyzing some of the behavior and properties of the signals and the system they represent. Deviation from nominal ranges of these parameters can be an indication of an anomaly \cite{hasnat22, hasnat22a}. In the power system context, the anomalies may indicate a \textit{stressed} system due to cyber attacks or physical events, such as line outages, generator trips, and abrupt load changes. In our previous works, local and global smoothness of bus voltage angle graph signals (i.e., $g_{\theta}$ and $l_{\theta}(n)$) have been utilized for the detection \cite{hasnat22}, location identification \cite{hasnat22}, characterization (including determining whether the stress is clustered or random, determining the stress center and radius) \cite{hasnat21}, and classification \cite{hasnat22b} of the stresses in the power system. The current work provides a focused study on the changing pattern of global and local smoothness values of different graph signals under single bus perturbation due to, for instance, abrupt changes in load demand or generation. Through this study, the spread of the effects of perturbation in the system will also be investigated through graph signal properties. Understanding the properties of stresses and their spread can support power system monitoring and planning, for instance, for predicting the grid instability due to load and generator changes, the effects of renewable energy resources on the system state, and for analyzing the effect of loads connected through grid-following and grid-forming inverters.

\subsection{Single Bus Perturbation}

A single bus perturbation $\mathscr{U}$ at the vertex (i.e., bus), $v_u \in \mathcal{S} \cup \mathcal{L}$ is defined by an abrupt change of value in the bus input at time $t_u$ and can be defined in the power graph signal form as:
\begin{equation}
    p(n, t_u) = p(n, t_u-\epsilon) + \Delta p_u(n),  
\end{equation}
where $\epsilon$ is a very small amount of time. The perturbation graph signal $\Delta p_u(n)$ can be modeled as a Kronecker Delta \cite{kundu12} graph signal:
\begin{equation}\label{defgamma}
    \Delta p_u(n) = \gamma \delta_u(n),  
\end{equation}
where $\delta_u(n)$ is the Kronecker delta graph signal defined as
$\delta_u(u)=1$ for $n = u$ and $\delta_u(n)=0$, for $n \neq u$ and $\gamma$ is a scalar called \textit{perturbation strength} associated with the perturbation, $\mathscr{U}$. Therefore, $\Delta p_u(u)=\gamma$. A positive value of $\gamma$ at the generator-only bus (i.e., $v_u \in \mathcal{S} \setminus \mathcal{L}$) indicates an increase in generated real power while a positive value of $\gamma$ at a load-only bus (i.e., $v_u \in \mathcal{L} \setminus \mathcal{S}$) indicates an increase in the real power load demand. The value of $\gamma$ in buses with both generators and loads (i.e., $v_u \in \mathcal{S} \cap \mathcal{L}$) can be described by the increase and decrease of both generations and loads. However, in  this work, only one change at a time (i.e., either an increase or decrease in generated power or load demand) is considered. It is also assumed that the inertia of the grid is negligible in response to the perturbation, $\mathscr{U}$. This assumption is reasonable for modern grids, where renewable energy resources are connected to the grid with inverters and loads are connected with converters.

In this work, the effects of the perturbation $\mathscr{U}$ on the voltage angle graph signal, $\theta(n)$ are evaluated. Let the \textit{\textbf{difference voltage angle graph signal}} due to the perturbation $\mathscr{U}$ at bus $u$ at time $t_u$ be defined as: 
\begin{equation}\label{defdeltheta}
    \Delta \theta_u(n)=|\theta(n,t_u)-\theta(n,t_u-\epsilon)|, 
\end{equation}
where $\epsilon$ is a small value. The signal values of the graph signal $\Delta \theta_u(n)$ have a direct relationship with the perturbation strength, $\gamma$. Therefore, for a better understanding of the dependency on the perturbation location, a normalized version of  $\Delta \theta_u(n)$ has been considered. The \textit{\textbf{normalized difference voltage angle graph signal}} is defined as:
\begin{equation} \label{defpsi}
    \psi_u(n) = \frac{\Delta \theta_u(n)}{\left| \gamma \right|}, 
\end{equation}
where $\psi(n)$ is expressed in $degree/mega-watt$. 

\begin{property}
Considering the DC Power flow model, the $\psi_u(n)$ depends only on the grid topology.
\end{property}

\textbf{\textit{Proof:}} Substituting the definition of $\theta(n)$ from equation (\ref{dcpf}) into equation (\ref{defdeltheta}):
\begin{equation}\label{delthetaresult}
    \begin{aligned}
& \Delta \theta_u(n)=\left| \sum_{j=1}^N \beta_{nj} p\left(j, t_u\right)-\sum_{j=1}^N {\beta}_{nj} p\left(i, t_u-\epsilon\right)\right| \\
& =\left|\sum_{j=1}^N \beta_{nj}\left[p\left(j,t_u\right)-p\left(j, t_u-\epsilon\right)\right]\right| \\
& =\left|\sum_{j=1}^N \beta_{nj} \Delta p_u(j)\right| \\
& =\left|\sum_{j=1}^N \beta_{nj} \gamma \delta_u(j)\right| \\
& =\left| \gamma \beta_{n u} \right|, \ \  (\text{using the property of Kronecker Delta \cite{kundu12}}). \\ 
&
\end{aligned}
\end{equation}

Next, substituting $\Delta \theta_u(n)$ into equation(\ref{defpsi}) leads to:
\begin{equation}
\label{psiBeta}
    \psi_u(n) = \frac{|\gamma \beta_{nu}|}{|\gamma|} = |\beta_{nu}|.
\end{equation}

This property shows that $\psi_u(n)$ does not depend on $\gamma$, under the DC power flow assumption. The normalized difference in voltage angle before and after the perturbation depends only on the location of the perturbation. In other words, the location of the perturbation affects $\psi_u(n)$ according to the topology of the grid, which captures the interconnections among the buses and the electrical distances between the components. Since the power system dynamics deviate from the DC power flow model, this property may not hold accurately in real power grids, nevertheless, it indicates that the effect of perturbation in the grid predominantly depends on its location rather than its strength. This property is important as it can be used for instance, for identifying the vulnerable buses with respect to perturbation issues, which is important for stability, maintenance, and resilience planning.   

The perturbation $\mathscr{U}$ affects the bus attributes of the perturbed bus, $v_u \in \mathcal{S} \cup \mathcal{L}$ as well as the other buses ($v \in \mathcal{V}, v \neq v_u$) in the system. The effects of the perturbation spread throughout the grid (similar to a stone causing ripples in the water). However, the effects are more complex in the power systems because of their irregular topology (i.e., non-Euclidean vertex domain) and complex interconnections based on the physics of electricity. While it is expected that the attributes of the nearby (geographical and topological) buses of the perturbed bus $v_u$ get affected more than the far-away buses, deviation from this expectation is very common. In other words, the relationship between the geographical/topological distance and the perturbation effects is irregular. In the next section, the spreadability of the perturbation $\mathscr{U}$ is studied in terms of the location of the perturbation and the perturbation strength.

\section{Effects of Single Bus Perturbation}

\subsection{Spreadability of Single Bus Perturbation}

For analyzing the spreadability of perturbation $\mathscr{U}$ in the grid in terms of the bus attributes, the changes introduced in the bus voltage angle graph signal at the buses at different hop distances from the perturbed bus, $v_u$ are evaluated. The mean of the signal values of $\psi_u(n)$ at all the vertices at $K-hop$ distance from the perturbed bus $v_u$ specifies how the buses at $K-hop$ distance are affected on average by the perturbation. This can be expressed as: 
\begin{equation}\label{defpsik}
    {\bar{\psi}}_u^{(K)} = \frac{1}{|{\mathcal{N}}_u^{(K)}|} \sum_{n \in {\mathcal{N}}_u^{(K)}} \psi_u(n), 
\end{equation}
where ${\mathcal{N}}_u^{(K)} \subset \mathcal{V}$ is the set of the $K-$ hop neighbors of $v_u$. According to Property 1, as $\psi_u(n)$ does not depend on the perturbation strength, ${\bar{\psi}}_u^{(K)}$ also does not depend on the perturbation strength under DC power flow assumptions.  

\begin{corollary}
 \textit{Under the DC Power flow assumption, the ${\bar{\psi}}_u^{(K)}$ depends only on the grid topology.}
\end{corollary}
 
\textbf{\textit{Proof:}}
Substituting $\psi_u(n)$ from equation (\ref{psiBeta}) into equation (\ref{defpsik}) results:
\begin{equation}\label{p1c1proof}
    {\bar{\psi}}_u^{(K)} = \frac{1}{|{\mathcal{N}}_u^{(K)}|} \sum_{n \in {\mathcal{N}}_u^{(K)}} |\beta_{nu}|. 
\end{equation}
Therefore, under DC power flow model ${\bar{\psi}}_u^{(K)}$ does not depend upon the perturbation strength, $\gamma$, rather depends upon the perturbation location, $v_u$. As such, ${\bar{\psi}}_u^{(K)}$ can be calculated from the susceptance matrix, $\mathbf{B}^{-1}$.  \qed

\begin{figure}[h]
    \centering
    \includegraphics[width=\columnwidth]{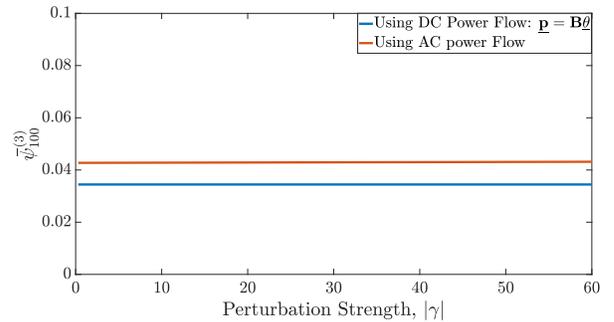}
    \caption{\small{Average of the values of normalized difference voltage angle graph signal at $3-$ hop distance from the perturbed bus $v_u=100$ for the IEEE 118 bus system \cite{118bus}.}}
    \label{p1c1}
\end{figure}

Fig. \ref{p1c1} shows ${\bar{\psi}}_{100}^{(3)}$, the average of the values of normalized difference voltage angle graph signal calculated from the equation ${\underline{\mathbf{\theta}}}=\mathbf{B^{-1}} \underline{\mathbf{p}}$ (DC power flow model) at $K = 3-$ hop distance from the perturbed bus no. $100$ of the IEEE 118 bus system \cite{118bus}. It can be observed that ${\bar{\psi}}_{100}^{(3)}$ is independent of the perturbation strength, $\gamma$. The results obtained from the AC power flow model in MATPOWER show a similar property, i.e., very weak dependence of ${\bar{\psi}}_{100}^{(3)}$ on $\gamma$. For the AC model results the value shows a slight variation (around 0.008 $degree/MW$) from the value obtained analytically using the DC power flow model.  

\begin{figure}[h]
    \centering
    \includegraphics[width=\columnwidth]{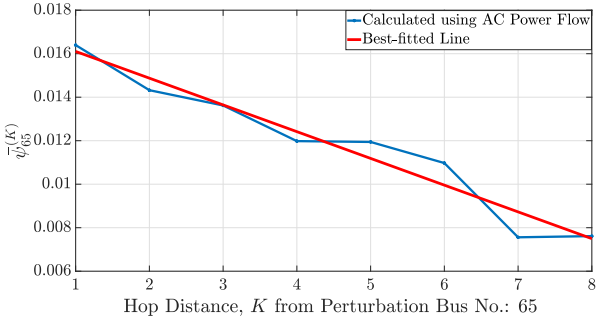}
    \caption{ \small{Average of the values of normalized difference voltage angle graph signal at different hop distances from the perturbed bus $v_u=65$ for the IEEE 118 bus system \cite{118bus}.}}
    \label{p1c12}
\end{figure}

The values of ${\bar{\psi}}_u^{(K)}$ show a decreasing trend as a function of $K$ as illustrated in Fig. \ref{p1c12} when calculated using the AC power flow model in MATPOWER. This behavior is expected as the effects of perturbation should spread and diminish from the source of the perturbation (i.e., $v_u$). Our experiments show that this decreasing trend is non-uniform over the grid and varies significantly depending on the location of the perturbation. Generally, a larger value of ${\bar{\psi}}_u^{(K)}$ at a far-away bus (i.e., a higher value of $K$) from the perturbation source indicates larger spreadability of the perturbation. Therefore, a flatter ${\bar{\psi}}_u^{(K)} \text{vs. } K$ curve indicates greater spreadability of the perturbation. As such, to quantify the spreadability, the slope of the best-fitted line (Fig. \ref{p1c12}, red straight line) to the ${\bar{\psi}}_u^{(K)} \text{vs. } K$ curve is defined as the \textit{spreadability measure}, $s$.  To this end, the spreadability measure due to the perturbation, $\mathscr{U}$ at bus $v_u$ can be expressed as:
\begin{equation}\label{defspreadability}
    s(u) = \frac{1}{\mathscr{S}[{\bar{\psi}}_u^{(1)} , {\bar{\psi}}_u^{(2)}, \dots {\bar{\psi}}_u^{(D)}]}, 
\end{equation}
where ${\mathscr{S}[{\bar{\psi}}_u^{(1)} , {\bar{\psi}}_u^{(2)}, \dots {\bar{\psi}}_u^{(D)}]}$ denotes the negative slope of best-fitted lines to the points $[{\bar{\psi}}_u^{(1)} , {\bar{\psi}}_u^{(2)}, \dots {\bar{\psi}}_u^{(D)}]$. 

\begin{corollary}
 \textit{Considering the DC Power flow model, $s(u)$ depends only on the location of perturbation, $\mathscr{U}$.}
\end{corollary}

\textbf{\textit{Proof:}} Since ${\bar{\psi}}_u^{(K)}$ is independent of $\gamma$ as proved in equation (\ref{p1c1proof}), from equation (\ref{defspreadability}), it can be shown that $s(u)$ is independent of $\gamma$ and only a function of the perturbation location $v_u$ under the DC power flow assumption.   \qed
 
\begin{figure}[h]
    \centering
    \includegraphics[width=\columnwidth]{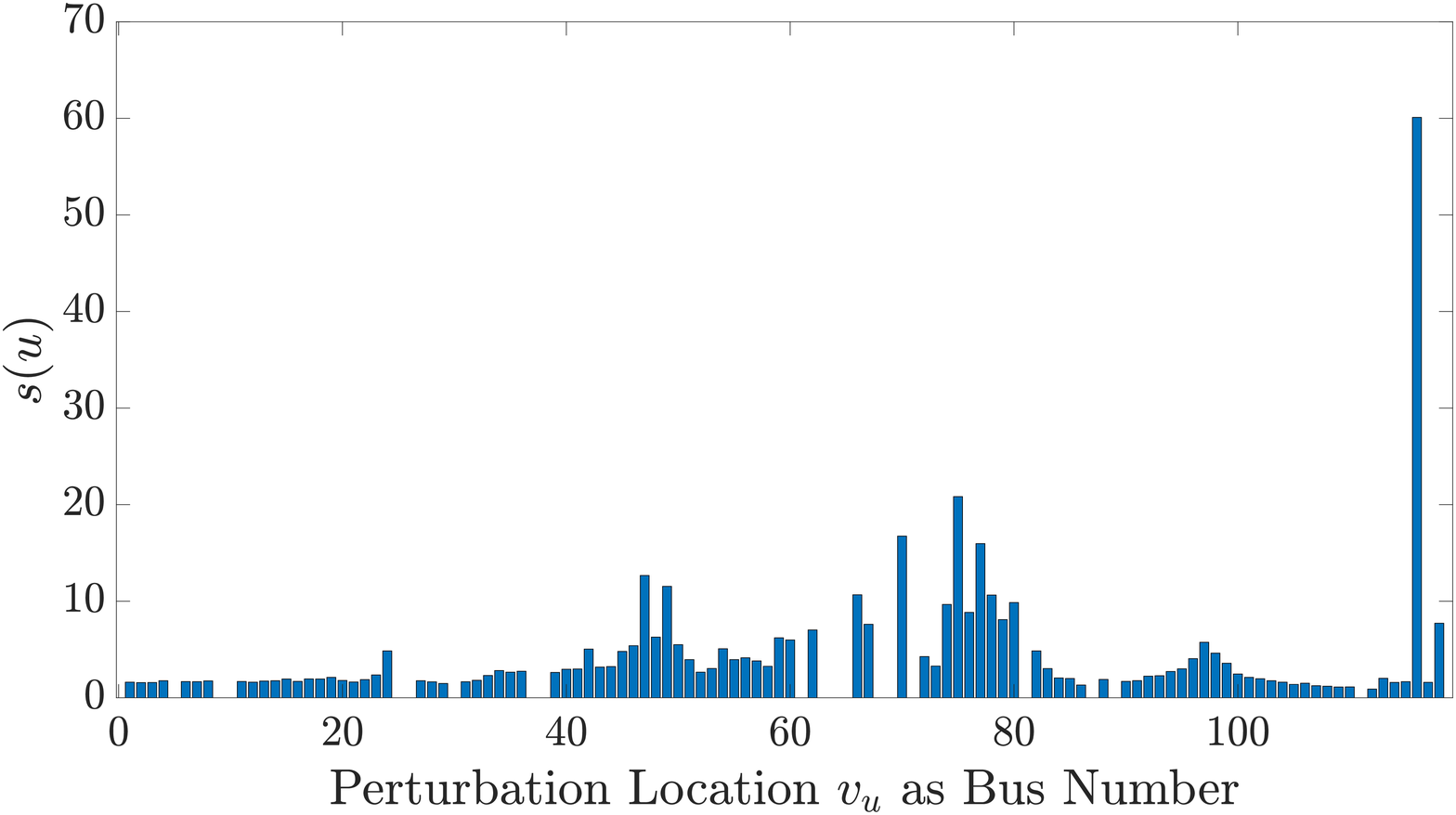}\\
    (a)
    \includegraphics[width=\columnwidth]{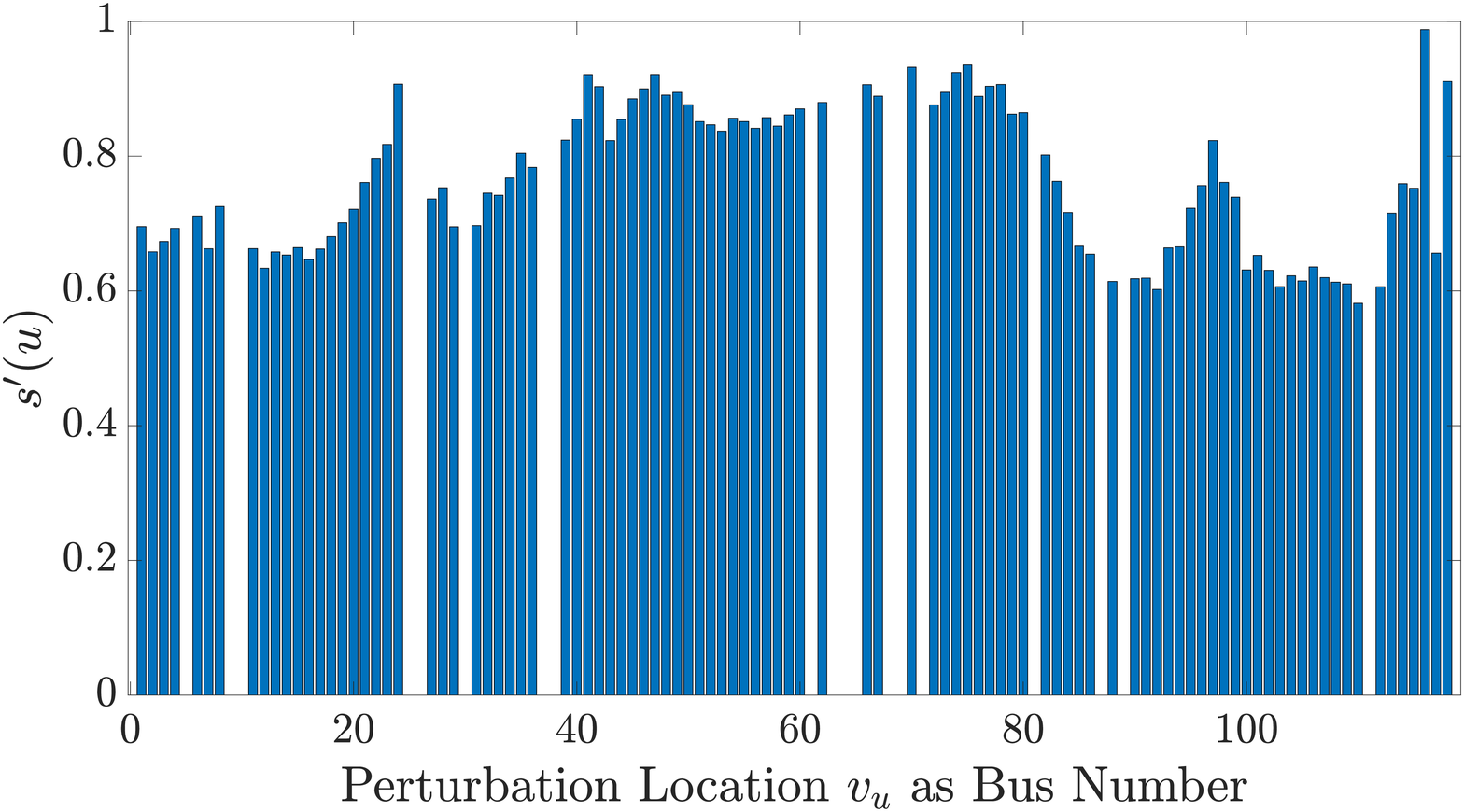}\\
    (b)
    \caption{\small{Spreadability values for a load perturbation of $50MW$ in different buses of the IEEE 118 bus system \cite{118bus}: (a) calculated using equation (\ref{defspreadability}), which is the proposed measure of spreadability in this paper, and (b) calculated using equation (\ref{defbuttner}), which is proposed in Buttner \textit{et. al.} \cite{buttner22}. The MATPOWER default loads and generations are considered as the pre-perturbation real power, $p(n,t_u-\epsilon)$. The similarity between the results in (a) and (b) from visual inspection might be challenging; however, their similarity can be understood by the Spearman rank correlation coefficient of $0.8562$.}}
    \label{p1c13}
\end{figure}

Fig. \ref{p1c13}(a) shows the spreadability measurement, $s(u)$ due to perturbation in different locations, $v_u \in \mathcal{L} \cup \mathcal{S}$ for a fixed perturbation strength. Based on the DC power flow model, the proposed spreadability measurement, $s(u)$ is shown to be unaffected by the perturbation strength. Meanwhile, the simulation results using the AC power flow model demonstrate a minimal dependence on the perturbation strength but a major dependence on the location of the perturbation. This observation indicates that the numerical findings align with the obtained theoretical results. Fig. \ref{p1c13}(a) illustrates how the effect of load perturbation in different load buses spreads through the grid as reflected in the difference bus voltage angle  graph signals. This observation can support the identification of vulnerable buses in the grid. These buses are susceptible to perturbations that can lead to more widespread effects and result in greater damage to the system. For example, from Fig. \ref{p1c13}(a) it is observable that the impact of a load perturbation at bus no. $116$ of the IEEE $118$ bus system is more spreadable through the grid compared to load perturbations at any other bus in the system. This result provides important insight, for instance, for maintenance and protection planning in the system. The presence of vulnerable areas with high spreadability indicates that these regions may not be suitable for the integration of renewable energy sources or EVs. Due to their susceptibility to perturbation spread, these areas may pose challenges for the reliable and stable operation of renewable energy and EV infrastructure \cite{sandhya23}.

Next, the introduced spreadability measure $s(u)$ in this work has been evaluated and compared with respect to the spreadability measure introduced in \cite{buttner22} based on a network-science-based approach. Here, the difference
voltage angle graph signal $\Delta \underline{\mathbf{\theta}}_u$ has been considered as the \textit{mean displacement vector} as defined in \cite{buttner22}. The spreadability measure introduced in \cite{buttner22}, is denoted as $s^{\prime}(u)$ and is defined as: 
\begin{equation}
    s^{\prime} (u) = C^{\prime}(u) \sum_{i=1}^N \frac{\Delta \theta_u(i)}{\sum_{j=1}^N \Delta \theta_u(j)} \mathscr{D} (v_u,v_i),
\end{equation}
where $\mathscr{D} (v_u,v_i)$ is the shortest path length between the perturbed bus $v_u$ and all the other buses $v_i \in \mathcal{V}$ in the graph $\mathcal{G}^{\prime}(\mathcal{V},\mathcal{E})$. This graph is defined by ignoring the weights of the graph $\mathcal{G}$ while having the same sets of vertices and edges. Moreover, the \textit{modified normalized closeness centrality}, denoted by $C^{\prime}(u)$ is defined  in \cite{buttner22} as: 
\begin{equation}\label{defsprime}
    C^{\prime}(n) = \frac{N}{\sum_{\forall v_i \in \mathcal{V}} \mathscr{D} (v_n,v_i)}.
\end{equation}

\begin{corollary}
 \textit{Considering the DC Power flow model, $s^{\prime}(u)$ is independent of the perturbation strength $\gamma$.}
\end{corollary}

\textbf{\textit{Proof:}} By substituting the expression of $\Delta \theta_u(n)$ from the equation (\ref{delthetaresult}) to the equation (\ref{defsprime}), it can be written that:
\begin{equation}\label{defbuttner}
    s^{\prime}(u) = C^{\prime}(u) \sum_{i=1}^N \frac{|\beta_{iu}|}{\sum_{j=1}^N |\beta_{ju}|} \mathscr{D}(v_u,v_i).
\end{equation}
The terms $C^{\prime}(u)$ and $\mathscr{D}(v_u,v_i)$ are calculated from the unweighted graph $\mathcal{G}^{\prime}$ for a certain perturbation location $v_u$ and therefore, depend only upon the interconnections among the buses of the grid. On the other hand, the $\beta_{ij}$ terms are related to the electrical parameters of the transmission line. Therefore, for a particular electrical grid,  $s^{\prime}(u)$ depends on the location of the perturbation and is independent of the perturbation strength.  \qed

Based on the results presented in Fig. \ref{p1c13}(a) and Fig. \ref{p1c13}(b), it can be observed that the proposed GSP-based spreadability measure $s(n)$ in this work and the network-science-based spreadability measure $s^{\prime}(u)$ from \cite{buttner22} show similarity for the $50MW$ of real power load perturbation at every bus of the system. The similarity of the results from these two measures is quantified by the \textit{Spearman's correlation coefficient} \cite{gauthier01} with the value $0.8562$ with no tied rank and a $p-value$ of $0$.   

In addition to evaluating the spreadability due to perturbations, it is important to evaluate other graph signal properties, which may be affected by the perturbation and may encode important information about the behavior of the system under perturbation. The global smoothness of graph signals describes the variation of values over buses in an aggregated form. Next, the effects of perturbation on the global smoothness of the bus voltage angle graph signals are discussed. In this analysis, load changes are considered as the main kind of perturbation.

\color{black}

\begin{property}
    \textit{Under the DC Power flow assumption, the global smoothness of the voltage angle graph signal is a quadratic function of the increased load.}
\end{property}

\textbf{\textit{Proof:}}
Let us start by writing the definition of the global smoothness for the voltage angle graph signal $\underline{\mathbf{\theta}}$ and use the DC power flow model to expand the definition of $\underline{\mathbf{\theta}}$ as follows: 
\begin{equation}\label{defgtheta}
    \begin{aligned}
g_\theta & =\frac{\underline{\mathbf{\theta}}^T \mathbf{L} \underline{\mathbf{\theta}}}{\underline{\mathbf{\theta}}^T \underline{\mathbf{\theta}}} \\
& =\frac{\left(\mathbf{B}^{-1} \underline{\mathbf{p}}\right)^{T} \mathbf{L} \left(\mathbf{B}^{-1} \underline{\mathbf{p}}\right)}{\left(\mathbf{B}^{-1} \underline{\mathbf{p}}\right)^{T}\left(\mathbf{B}^{-1} \underline{\mathbf{p}}\right)} \\
& =\frac{\underline{\mathbf{p}}^{T}\left(\mathbf{B}^{-1}\right)^{T} \mathbf{L} \mathbf{B}^{-1} \underline{\mathbf{p}}}{\underline{\mathbf{p}}^{T}\left(\mathbf{B}^{-1}\right)^{T} \mathbf{B}^{-1} \underline{\mathbf{p}}} \\
& =\frac{\underline{\mathbf{p}}^{T} \mathbf{Q} \underline{\mathbf{p}}}{\underline{\mathbf{p}}^{T} \mathbf{R} \underline{\mathbf{p}}}
\end{aligned}
\end{equation}
Here, $\mathbf{Q}=\left(\mathbf{B}^{-1}\right)^{T} \mathbf{L} \mathbf{B}^{-1}$ and $\mathbf{R}=\left(\mathbf{B}^{-1}\right)^{T} \mathbf{B}^{-1}$, both contain topological information and are independent of $\underline{\mathbf{p}}$. Since the other elements of the vector $\underline{\mathbf{p}}$, except the $u-$th element, are the same before and after the perturbation, $\mathscr{U}$ (as described in Section III), $g_\theta$ is a \textit{quadratic function} of the real power $p(u,t_u)$ at the perturbed bus $v_u \in \mathcal{V}$.  Specifically, from equation (4) and equation (5), the global smoothness can be written as:
\begin{equation}
    \begin{aligned}
& g_\theta \propto p^2\left(u, t_u\right) \\
& \begin{aligned}
\Rightarrow g_\theta \propto[
p^2\left(u, t_u-\epsilon\right)
 + \gamma^2
+2 \gamma p\left(u, t_u-\epsilon\right)]
\end{aligned} \\
& \Rightarrow g_\theta \propto \gamma^2+2 \gamma p\left(u, t_u-\epsilon\right)
&
\end{aligned}
\end{equation}

\begin{figure}[h]
    \centering
    \includegraphics[width=\columnwidth]{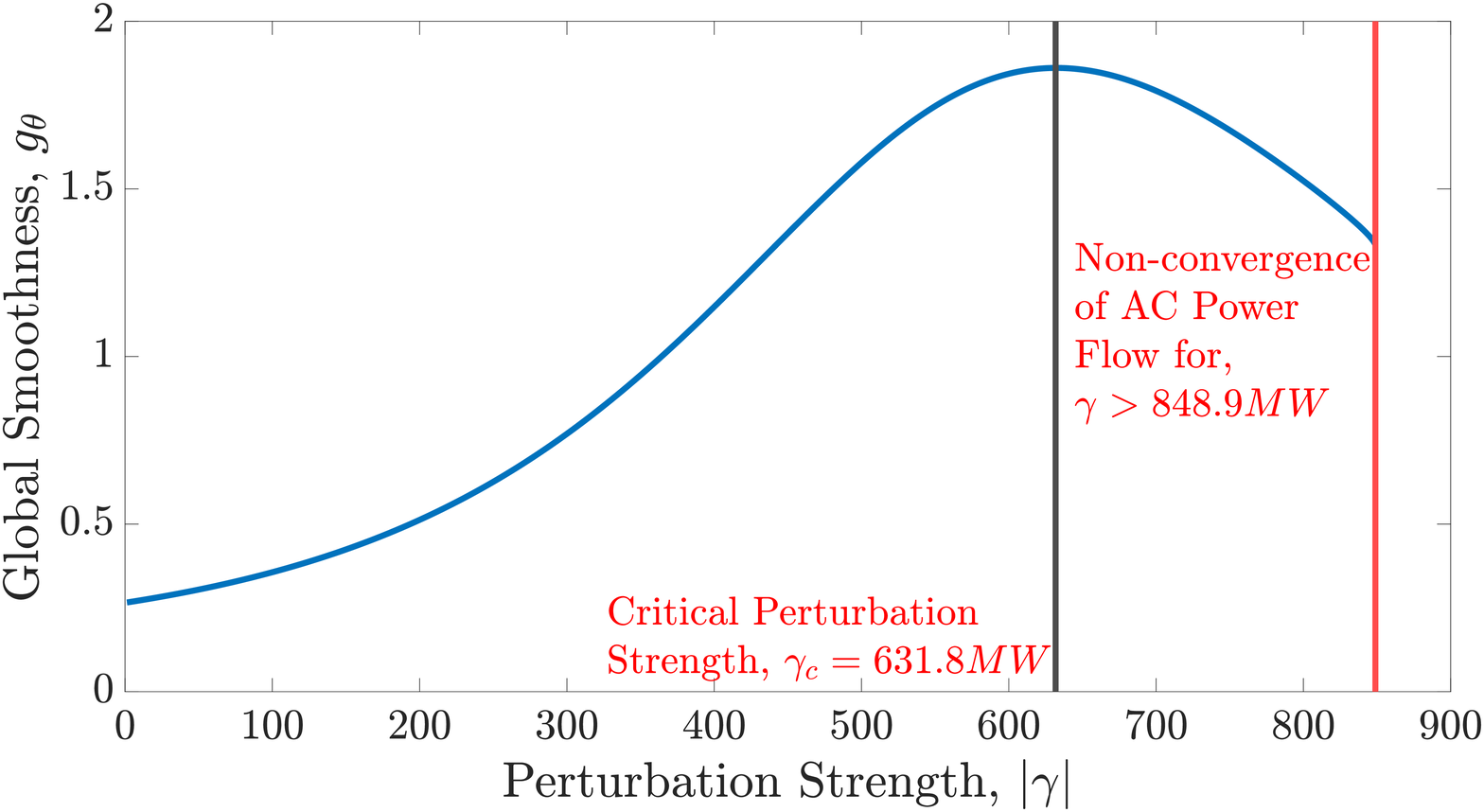}
    \caption{\small{Global smoothness of the bus voltage angle graph signal as a function of perturbation strength in the case of real-power load perturbation at bus $102$ of the IEEE $118$ bus system calculated using AC power flow in MATPOWER. At the critical perturbation, $\gamma_c$ the global smoothness is maximum and begins to drop beyond this point. The numerical evaluation showed that any increase of perturbation strength beyond $\gamma_{nc}$ leads to the non-convergence issue in the power flow calculations.}}
    \label{p2a1}
\end{figure}

Therefore, $g_\theta$ is a quadratic function of $\gamma$. Fig. \ref{p2a1} shows $g_{\theta}$ as a function of perturbation strength $\gamma$ for load perturbation at bus no. $16$ of the 118 IEEE bus system. The values of $g_\theta$ are calculated using equation (\ref{defg}) with the values of $\theta(n)$ obtained from the AC power flow model in MATPOWER. Although Property~2 is derived under the DC power flow assumption, Fig. \ref{p2a1} shows that it also holds for the AC power flow (although with some numerical deviation). The quadratic form of $g_\theta$ as the function of perturbation strength can have important implications. For instance, our experiments have shown that an increasing trend in $g_\theta$ may indicate a stressed system. Specifically, the power grid bus voltage angle graph signal is generally smooth over the vertices under normal operating conditions \cite{hasnat22,ramakrishna21}. Therefore the value of $g_\theta$ generally stay small, while the actual value depends on several factors, such as the system topology, load demand, and generation amount in the system. From Fig. \ref{p2a1} it can be observed that when the load is increasing continually at a particular bus, initially $g_{\theta}$ increases with the increasing load, which indicates increasing fluctuations of signal values from vertex-to-vertex until the perturbation strength reaches a critical point $\gamma_{c}$ (associated with a critical load demand of $p_{d_{c}}(u)$) for the  perturbed bus, $v_u$. Increasing the load beyond this critical point results in decreasing values of $g_{\theta}$, which in general can indicate smoother signal and normal grid conditions. However, in this particular case, the decrease in the global smoothness after reaching its maximum suggests a stressed system, and the issue of non-convergence of the AC power flow calculations rise in this phase.

Moreover, the increase of, $p_{d}(u)$, i.e., $\gamma$, at the perturbed bus increases the power flow through a number of transmission lines. Increasing the size of perturbation can lead to overloading of transmission lines and outages and in severe cases cascading failures.

\begin{application}
    Determining the critical value of perturbation strength, $\gamma$ for smooth grid operation. 
\end{application}

The critical value of the perturbation strength, $\gamma$, which also corresponds to the critical load size at bus $v_u$ can be identified based on the maximum values of $g_\theta$ as follows: 
\begin{equation}\label{eqextrema}
  \frac{\partial g_\theta}{\partial p(u)} \Big\vert_{p_{d}(u)=p_{d_c}(u)}= \frac{\partial g_\theta}{\partial p(u)} \Big\vert_{\gamma=\gamma_c} = 0
\end{equation}

By substituting equation (\ref{defgtheta}) into equation (\ref{eqextrema}) and applying the rules of matrix differentiation:

\begin{equation}
    \underline{\mathbf{p}}^T \mathbf{Q} \underline{\mathbf{p}} \frac{\partial g_{\theta}}{\partial p(u)} (\underline{\mathbf{p}}^T \mathbf{R} \underline{\mathbf{p}}) = \underline{\mathbf{p}}^T \mathbf{R} \underline{\mathbf{p}} \frac{\partial g_{\theta}}{\partial p(u)} (\underline{\mathbf{p}}^T \mathbf{Q} \underline{\mathbf{p}})
\end{equation}

By solving the equation for $p(u)$ which is the same as the $u-$th element of $\underline{\mathbf{p}}$ the value of real power for which $g_\theta$ is maximum can be obtained, and therefore the critical perturbation strength $\gamma_c$ can be obtained by equation (\ref{defgamma}).

Fig. \ref{p2a1} shows $g_\theta$ for monotonous load increase at bus $17$ of the IEEE 118 bus system \cite{118bus} (which is purely a load bus). The result presented in this figure suggests that the perturbation strength of $\gamma_c = 631.8 MW$ results in the maximum $g_\theta$ value and corresponds to our defined critical load. This critical load advises on a stressed system for which the power flow non-convergence based on the numerical results occurred at the perturbation strength of $\gamma_{nc} = 848.9$ corresponding to a load size of $853.9MW$.

\begin{property}
    \textit{Under the DC Power flow assumption the global smoothness of the difference voltage angle graph signal, $\Delta \theta$ is independent of the perturbation strength.}
\end{property}

\textbf{\textit{Proof:}}
Following the definition of global smoothness in equation (2), the global smoothness of the difference bus voltage angle graph signal $\Delta \theta_u(n)$ before and after the perturbation $\mathscr{U}$ can be written as:
\begin{equation}
g_{\Delta \theta}=\frac{\sum_{i=1}^N \sum_{j=1}^N L_{i j} \Delta \theta_u(i) \Delta \theta_u(j)}{\sum_{k=1}^N \Delta \theta_u^2(k)}. \\
\end{equation}
By substituting $\Delta \theta_u(n)$ from the result expressed in equation (\ref{delthetaresult}), it can be written that:

\begin{equation}
\begin{aligned}
& g_{\Delta \theta} =\frac{\sum_{i=1}^N \sum_{j=1}^N L_{i j}\left|\gamma \beta_{i u}\right|\left|\gamma \beta_{j u}\right|}{\sum_{k=1}^N\left|\gamma \beta_{k u}\right|\left|\gamma \beta_{k u}\right|} \\
& =\frac{\sum_{i=1}^N \sum_{j=1}^N L_{i j}\left|\beta_{i u} \beta_{j u}\right|}{\sum_{k=1}^N\left|\beta_{k u}\right|^2}.
&
\end{aligned}
\end{equation}

\begin{figure}[h]
    \centering
        \includegraphics[width=\columnwidth]{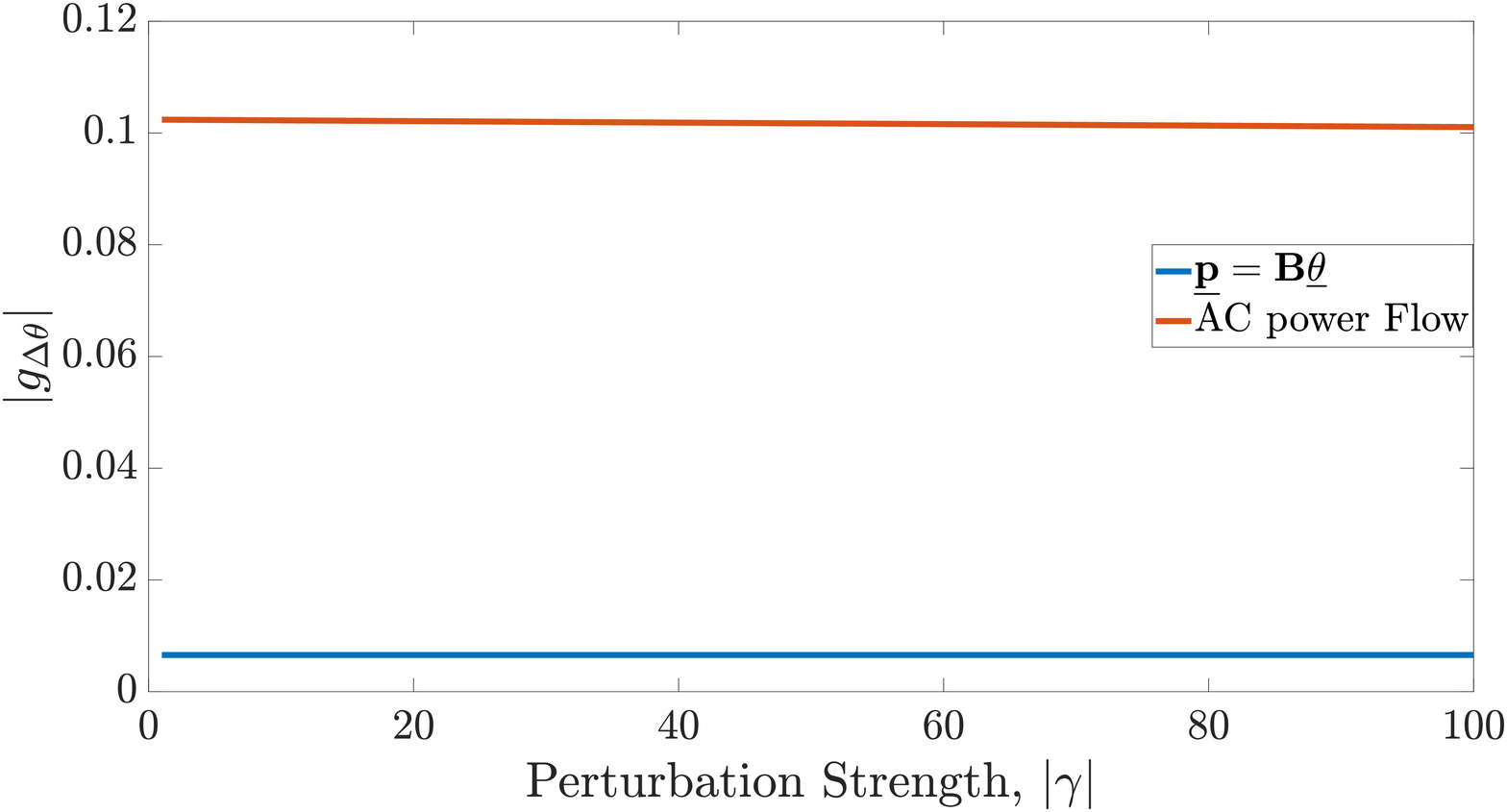}\\
        (a)\\
        \includegraphics[width=\columnwidth]{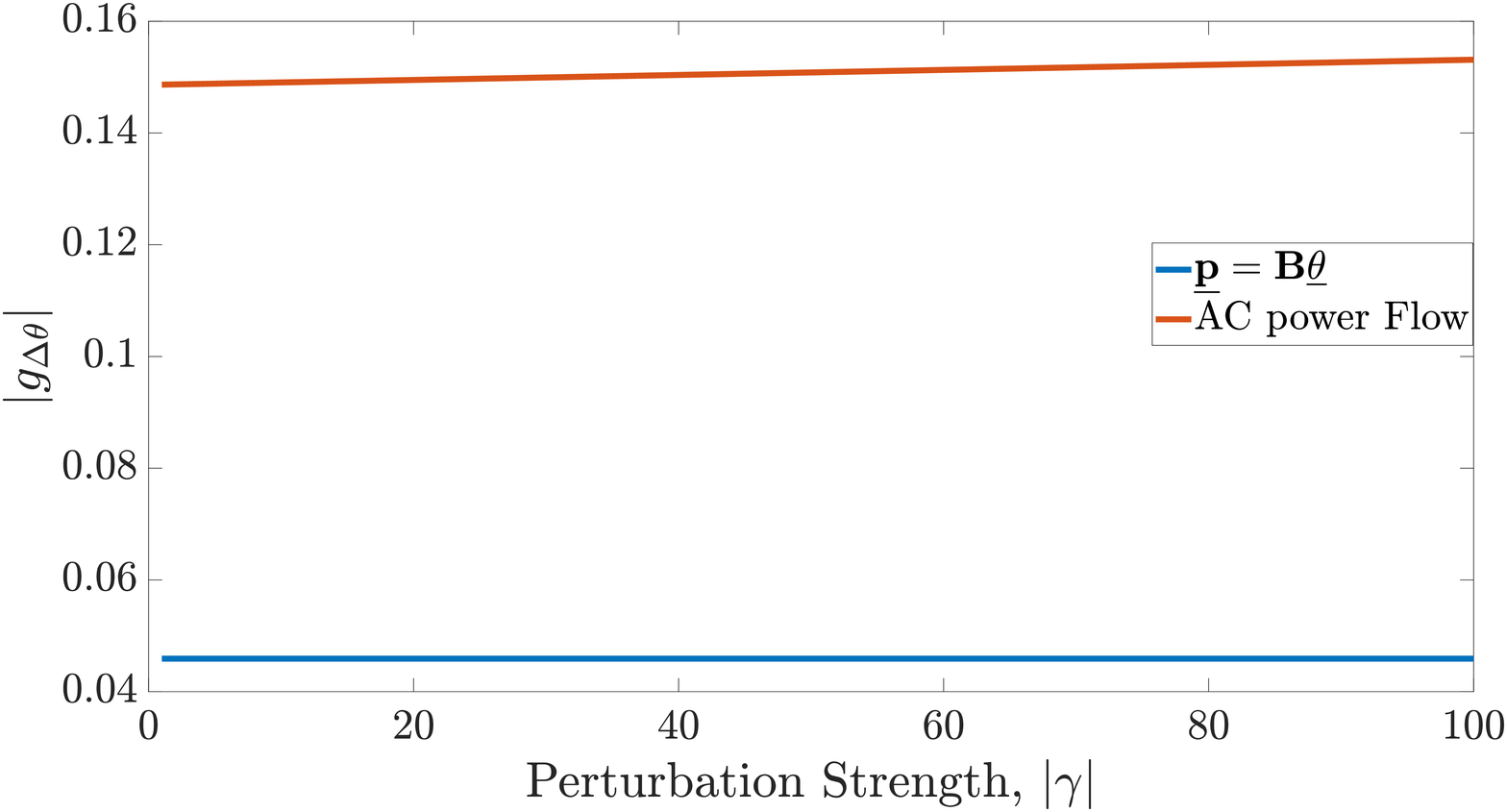}\\
        (b)\\
    \caption{\small{The absolute value of global smoothness of the difference bus voltage angle graph signal before and after the perturbation as a function of perturbation strength $\gamma$ for two sample locations as (a) load perturbation in Bus no. $8$ and, (b) generation perturbation in Bus no. $10$ of the IEEE 118 bus system \cite{118bus}. The results show that $|g_{\Delta\theta}|$ is independent of strength for the DC power assumption, based on the power flow calculation using $\mathbf{\underline{p}} = \mathbf{B} \mathbf{\underline{\theta}}$ and shows a very minor dependency on $\gamma$ for the AC power flow results in both perturbation cases.}}
    \label{p3}
\end{figure}

Since there is no $\gamma$ present in the right-hand side of the equation, $g_{\Delta \theta}$ does not depend on the perturbation strength, but rather depends on the topology of the system. \qed

This GSP-based property associated with both real power load perturbation (Fig. \ref{p3}(a)) and real power generation perturbation (Fig. \ref{p3}(b)) has been evaluated by simulations on the IEEE $118$ bus system. From Fig. \ref{p3}, it can be observed that in perturbations in both cases, power flow calculation using $\mathbf{\underline{p}} = \mathbf{B} \mathbf{\underline{\theta}}$ under DC power flow yields a constant function for $|g_{\Delta\theta}|$ vs. $\gamma$, which indicates the independence on the perturbation strength. The AC power flow results also justify this property, while showing a minor dependency on the perturbation strength.

\begin{figure}[h]
    \centering
    \includegraphics[width=\columnwidth]{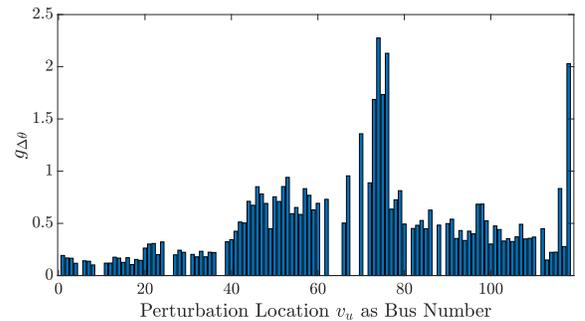}
    \caption{\small{Global smoothness of the difference bus voltage angle graph signal before and after the perturbation for the same amount of load perturbation $\gamma=50MW$ in each of the load buses of the IEEE $118$ bus system.}}
    \label{p32}
\end{figure}

This property enables $g_{\Delta\theta}$ to be a GSP-based measure for evaluating the effects of perturbation in different locations in the system. Fig. \ref{p32} shows the values for $g_{\Delta\theta}$ for a perturbation of $\gamma=50MW$ in each of the load buses of the IEEE $118$ bus system. From this result, it can be observed that load perturbations of the same strength at different buses have different effects in the grid, which is reflected on the graph signal $\Delta\theta(n)$ and its smoothness. Note that the graph signal $\Delta\theta(n)$ (being a difference graph signal before and after the perturbation) inherently contains some time evolution information and can help  characterize the spread patterns of perturbations. This can be understood from the visual resemblance of the bar diagram of  $g_{\Delta\theta}$ in Fig. \ref{p32} with the bar diagram of our proposed spreadability measure, $s(u)$ in Fig. \ref{p1c13}. The similarity between $g_{\Delta\theta}$ and $s(u)$ can be also justified by the \textit{cosine similarity} \cite{xia2015} of $0.8281$ and \textit{Spearman rank correlation co-efficient} \cite{gauthier01} of $0.61$ for $\gamma=50MW$ perturbations in all the load buses of the IEEE $118$ bus system. Therefore, the GSP-based parameter $g_{\Delta\theta}$ highlights the reliance of perturbations on locations within the grid, particularly in assessing the extent to which the effects of the perturbation can spread.
Similar results can be observed in the local smoothness of the graph signal $\Delta\theta_u(n)$.

\begin{property}
    \textit{Under the DC Power flow assumption, the local smoothness of the difference voltage angle graph signal, $\Delta \theta$ is independent of the perturbation strength.}
\end{property}

\textbf{\textit{Proof:}}
From equation (\ref{deflocsmooth}), the local smoothness at bus $n$ for the graph signal $\Delta \theta_u(n)$ (which is the difference bus voltage angle graph signal before and after the perturbation $\mathscr{U}$), can be calculated as:
\begin{equation}
    l_{\Delta \theta}(n)=\frac{\sum_{k=1}^N L_{nk} \Delta \theta_u(k)}{\Delta \theta_u(n)}, \Delta \theta_u(n) \neq 0. \\
\end{equation}
Substituting the $\Delta \theta_u(n)$ from equation (\ref{delthetaresult}) in the above equation results in:
\begin{equation}\label{ldelthetau}
    l_{\Delta \theta}(n)= \frac{\sum_{k=1}^N L_{nk} |\gamma\beta_{k u}|}{|\gamma \beta_{n u}|} = \frac{\sum_{k=1}^N L_{nk} |\beta_{k u}|}{|\beta_{n u}|}, \ \ \Delta \theta_u(n) \neq 0. \\
\end{equation}
Equation (\ref{ldelthetau}) provides the local smoothness values of $\Delta\theta_u(n)$ at every vertex, $v_n$ of the graph. The local smoothness values at the perturbed bus can be obtained by putting $n=u$ in equation (\ref{ldelthetau}) as: 
\begin{equation}
    l_{\Delta \theta}(u)=\frac{\sum_{k=1}^N L_{u k} |\beta_{k u}|}{|\beta_{u u}|} ,  \beta_{u u} \neq 0, \\
\end{equation}
which is independent of the perturbation strength. \qed

\begin{figure}[h]
    \centering
    \includegraphics[width=\columnwidth]{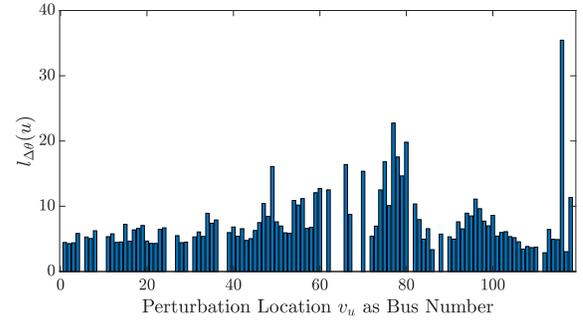}
    \caption{\small{Local smoothness values at the perturbation vertex of the difference bus voltage angle graph signal before and after the perturbation for the same amount of load perturbation $\gamma=50MW$ in each of the load buses of the IEEE $118$ bus system.}}
    \label{p4}
\end{figure}

The lack of dependence on perturbation strength makes $l_{\Delta \theta}(u)$ a suitable measure for analyzing the locational dependence of perturbations in the grid, similar to $g_{\Delta \theta}$. Like $g_{\Delta \theta}$, the local smoothness value of the difference bus voltage angle graph signal before and after the perturbation, assessed at the perturbation point, can be utilized as an estimator of the spreadability of the perturbation effect. Fig. \ref{p4} shows the values of local smoothness at the perturbed vertices due to the same amount of load perturbation $\gamma=50MW$ at each load bus of the IEEE $118$ bus system. The bar diagram of $l_{\Delta \theta}(u)$ seems similar to the bar diagram of our proposed spreadability measure, $s(u)$ for IEEE $118$ bus system. The cosine similarity \cite{xia2015} and the Spearman rank correlation coefficient \cite{gauthier01} between $s(u)$ and $l_{\Delta \theta}(u)$ for $50MW$ perturbations are, respectively, $0.8925$ and $0.66$, which suggests that $l_{\Delta \theta}(u)$ can serve as a GSP-based estimator of perturbation spread. 

\section{Conclusion}
This article presents a perspective based on GSP regarding the impacts of a single bus perturbation in the electrical grid. The perturbation is characterized by a sudden change in the real-power load demand or generation. Specifically, the article investigates the effects of the perturbation by considering its spread throughout the grid. A measure of spreadability based on GSP is proposed, and it is demonstrated that both global and local smoothness measures of the difference bus voltage angle graph signal can be used as estimators of the spreadability of the perturbation.  The findings indicate that the proposed measure of spreadability, along with the local and global smoothness properties of the graph signals, are not influenced by the perturbation strength. Instead, these properties primarily depend on the location of the perturbation. Furthermore, the article characterizes the global smoothness of the bus voltage angle graph signal as a quadratic function of the perturbation strength. It is shown that beyond a critical perturbation strength, the global smoothness starts to decrease, and further increases in perturbation strength may result in power flow divergence, which can be indicative of a stressed system.

The present study builds upon the DC power flow model assumption and employs a simple and generic perturbation model. Nevertheless, this research offers intriguing insights into the impact of perturbations in the grid and introduces a new perspective on utilizing GSP for analyzing various problems in power systems, for instance, cascading failures, as perturbation analysis. For example, such analyses can help characterize whether a perturbation can create a cascade or define how the failures propagate relative to the location, strength, and nature of the perturbation.

\section*{Acknowledgment}

This material is based upon work supported by the National Science Foundation under Grant No.~2238658.

\end{document}